# Simulations of surface charge density changes during the untreated solid tumor growth

Henry Bory Prevez[1], Argenis Adrian Soutelo Jimenez[2,†], Eduardo José Roca Oria[3], José Alejandro Heredia Kindelán[3], Maraelys Morales González[4], Narciso Antonio Villar Goris[5,6], Nibaldo Hernández Mesa[7], Victoriano Gustavo Sierra González[8], Yenia Infantes Frometa[9], Juan Ignacio Montijano Torcal[10], Luis Enrique Bergues Cabrales[5,10,*]

[1]Departamento de Control Automático, Facultad de Ingeniería Eléctrica, Universidad de Oriente, Santiago de Cuba, Cuba
[2]Departamento de Química, Facultad de Ciencias Naturales y Exactas, Universidad de Oriente, Santiago de Cuba, Cuba
[3]Departamento de Física, Facultad de Ciencias Naturales y Exactas, Universidad de Oriente, Santiago de Cuba, Cuba
[4]Departamento de Farmacia, Facultad de Ciencias Naturales y Exactas, Universidad de Oriente, Santiago de Cuba, Cuba
[5]Departamento de Ciencia e Innovación, Centro Nacional de Electromagnetismo Aplicado, Universidad de Oriente, Santiago de Cuba, Cuba
[6]Universidad Autónoma de Santo Domingo, Santo Domingo, República Dominicana
[7]Centro de Neurociencias de Cuba (CNEURO), La Habana, Cuba.
[8]Grupo de las Industrias Biotecnológica y Farmacéuticas (BioCubaFarma), La Habana, Cuba
[9]Facultad de Ciencias Sociales, Universidad de Oriente, Santiago de Cuba, Cuba
[10]Departamento de Matemática Aplicada, Instituto Universitario de Matemática y Aplicaciones, Universidad de Zaragoza, Zaragoza, España.

## Abstract

Understanding the untreated tumor growth kinetics and its intrinsic findings is interesting and intriguing. The aim of this study is to propose an approximate analytical expression that allows to simulate changes in surface charge density changes at cancer-surrounding healthy tissue interface during the untreated solid tumor growth. For this, the Gompertz and Poisson equations are used. Simulations reveal that the unperturbed solid tumor growth is closely related to changes in the surface charge density over time between the tumor and the surrounding healthy tissue. Furthermore, the unperturbed solid tumor growth is governed by temporal changes in this surface charge density. It is concluded that graphic strategies corroborate the correspondence between the electrical and physiological parameters in the untreated cancer, which may have an essential role in its growth, progression, metastasis and protection against immune system attack and anti-cancer therapies. In addition, the knowledge of surface charge density changes at cancer-surrounding healthy tissue interface may be relevant when redesigning the molecules in chemotherapy and immunotherapy taking into account their polarities. This can also be true in the design of completely novel therapies.

---

† argenis@uo.edu.cu
* Corresponding author: berguesc@yahoo.com

# Surface charge density changes in tumor growth

## 1    Introduction

The untreated solid tumor growth kinetics (TGK) exhibits a sigmoidal shape and its understanding constitutes a challenge for researchers [1,2]. Several equations are used to describe TGK, such as: the conventional Gompertz (CGE, the most accepted) [1,3], Montijano-Bergues-Bory-Gompertz [4], modified Kolmogorov-Johnson-Mehl-Avrami, Logistic and Bertalanffy [1]. The parameters of three first equations are interconnected [5]. These equations involve different biological parameters (i.e., intrinsic growth rate, endogenous anti-angiogenesis, carrying capacity of the tumor) obtained from fitting experimental data (tumor mass and volume, and cancer cells number). Nevertheless, none of these equations include bioelectrical parameters related to electrical properties and the active bioelectricity (or bioelectric potential) of the cancer, surrounding healthy tissue, and interface between these two tissues, named $\Sigma$.

The electrical properties and active bioelectricity inherent in cancer and surrounding healthy tissue are experimentally confirmed by means of several techniques, such as: bioelectrical impedance analysis [6], image technique of the electric current density [7], microelectrodes and neutralized input capacity amplifiers, high-impedance micropipettes, potentiometry, fluorescence, electrical double layer in field-effect transistors, electrical impedance spectroscopy together with other devices [8-14]. Additionally, vibrating probes, glass microelectrodes, microfluidic-based tissue/organ-on-a-chip devices, and endoscopes with inserted electronics to detect bioelectricity changes in real-time are recommended. Nanotechnology-based bioelectronics with nano-sized devices is used to quickly detect cancer at an earlier stage [8,9]. The bioelectricity-driven nanoparticle binding is suggested instead of static electrical potential via electrophoresis. The bioelectricity is proposed to capture electrostatically and magnetically circulating cancer cells from the entire blood to investigate the metabolic state of them [15].

Several findings have been revealed, such as: 1) differences between electrical conductivities ($\eta_k$, k = 1,2) and electrical permittivities ($\varepsilon_k$, k = 1,2) of the untreated malignant tumor (k = 1) and the surrounding healthy tissue (k = 2) [11]; 2) these two physical properties as a potential diagnostic method [16]; 3) differences between ionic current (due to the movement of charged ions) and faradic current (produced by electrons exchange from reduction and/or oxidation of biochemical molecules) in cancer and surrounding healthy tissue [8]; 4) the existence of chemical and electrical (charged negatively) environments in cancer cells and untreated tumors [1,4,14] and their key roles in the genesis, growth, progression, metastasis and treatment of cancer [17]; 5) the impact of tumor microenvironment on its electrical properties [16]; 6) the breakdown of intercellular communication (gap junction) in the tumor due to low regulation in expression of the connexin [12,18]; 7) negative electrical biopotentials in the tumor and positive electrical biopotentials in the surrounding healthy tissue [12-14]; 8) cancer cells and some cells of the immune system negatively charged [14]; 9) weak the electrical coupling among cancer cells and the association of deregulation of intercellular communication with tumorigenicity and metastasis of the cancer [14,18]; 10) bioelectronic cancer regulator as an initiator of the mitosis and deoxyribonucleic acid synthesis; and 11) correction of alterations in the electrical communication system of cancer by manipulating its bioelectrical properties, named bioelectronic medicine [8].

The uneven movement of ions and electrons across the plasma membrane via ion pumps modify the imbalance of the charge between the intra- and extracellular compartments. This ionic imbalance, gene expression level, glutamate-dependent currents, and both ionic and faradic currents explain the active cancer bioelectricity [8]. Furthermore, the ionic imbalance on both sides of the cancer cell membrane involved in the deregulation of ionic activity (a novel hallmark of cancer cells), altered membrane electrical potential difference ($V_{mem}$), shape change, pH, heterogeneity, phenotype, metabolism abnormalities, growth signaling, proliferation, tumorigenesis, angiogenesis, invasion and metastasis of the cancer cells, as well as in the plasticity, heterogeneity and cellular networks of cancer [8,16,19-22]. The intra-tumor heterogeneity and anisotropy have an essential role in its growth, metastasis and resistance to anti-cancer therapies [1,2,4]. The cancer phenotypes include both cellular ionic and faradic currents. The tumor growth may be due to malfunctions in bioelectrical circuitry of their cells. The tumor progression may be explained by the alterations of trans-plasma membrane electron transport. And the tumor metastasis considers the degradation of basement membranes, cancer cell invasion, migration, extravasation, and colonization [8].

Biological processes form bioelectric circuits from individual cell behaviors and anatomical information encoded in bioelectrical states to achieve a better control over spatiotemporal biological patterns. Electrically active cancer cells possess bioelectric circuitry that generates resting membrane potential and endogenous electric fields that influence cell functions and communication [8,23]. Endogenous electric potential gradients (established across multiple cells due to gap junctions and other cell-to-cell connections on a tissue level) induce small endogenous electric fields, which are responsible for altered migration and invasiveness of cancer cells [18,24].

Alterations in $V_{mem}$ are involved in high proliferation (due to the depolarization of their membranes by higher intracellular concentration of sodium ions) and mitosis, depletion of adenosine triphosphate, fail of ionic pumps at the cellular membrane and mechanism of contact inhibition of cancer cells. A depolarized membrane is considered a driving force for the production of $Ca^{2+}$ and bioelectronic cancer regulator that affect proliferation, migration, invasion and metastasis of cancer cells [8,12-14]. Furthermore, changes in $V_{mem}$ are related to the modulation of local concentrations of signaling molecules and ions, the spatiotemporal regulation of morphogenesis, the interaction with heterogeneous networks (that combines conventional gene regulatory network) is controlled by spatiotemporal bioelectrical patterns based on electric potentials and currents from steady and oscillatory multicellular states, among others. In turn, these spatiotemporal bioelectrical patterns influence on the spatiotemporal distributions of signaling ions and molecules that modulate biochemical pathways in cancer cells, and therefore in growth and regeneration [8,9,25].

$V_{mem}$ may be regulated in different ways, such as: the ion channel expression, the ionic composition of the extracellular environment, and the presence of bioelectronic gradients within cancer [8,20]. Therefore, Payne et al. [20] suggest that $V_{mem}$ should be analyzed in two directions: $V_{mem}$ effect on the cellular function (that contributes to the cancer phenotype) and how $V_{mem}$ is affected (by the expression of voltage-gated ion channels and cell metabolism). Alterations in the metabolism of cancer cells modify $V_{mem}$ [8,15,26]. Bioelectrical pathways associated with a metabolic phenomenon affect ionic electrical-based communication among cancer cells, like: reactive oxygen species and aberrant trans plasma membrane electron transport systems. The reverse Warburg effect is induced in cancer cells by higher levels of reactive oxygen species, which may be caused by malfunction in the redox balance, altered biological electron transfer reactions (higher electron transfer), a high energetic demand, increased concentration of reduced bioelectrochemical mediators, and participation of the trans plasma membrane electron transport systems in oxidation and redox centers existing in cell membranes transport electrons across these membranes in the form of faradic currents [8,26].

The results above-mentioned corroborate the close relationship between biological and electrical parameters in tissues. Likewise, it confirms that the bioelectrical activity in all cell types (i.e., cancer) is involved in many physiological mechanisms. Nevertheless, bioelectrical pathways are still poorly understood in cancer cells, TGK and $\Sigma$. This should be taken into consideration because both tissue types have different electrical properties and bioelectrical activities. Therefore, understanding of the bioelectricity in cancer and surrounding healthy tissue constitutes a challenge for researchers.

It is documented in Electrodynamics of media that a surface charge density arises at the interface between two materials in contact with different electrical properties [27]. Therefore, a surface charge density ($\sigma_{12}$) is expected at $\Sigma$ for the following reasons: firstly, solid tumors have chemical and electrical environments [1,4,14]. Secondly, the cancer and its surrounding healthy tissue are in contact and heterogeneous [28]. Thirdly, both tissue types differ significantly in their electrical properties and thermal [10,13,14,29,30], and physiological parameters [8,14]. Fourthly, $\sigma_{12}$ is due to synergism between an external volumetric current density (the source of electricity) and the Maxwell-Wagner-Sillars interfacial polarization. The Maxwell-Wagner-Sillars effect is an interfacial relaxation process that occurs for all two-phase multi-systems, in which the electric current must pass an interface between two different loss dielectrics [28,31,32]. Lastly, the electrophysiological activity in cancer (in tumor regions near $\Sigma$ mainly) is higher than that in the surrounding healthy tissue [8,9,14,33].

$\sigma_{12}$ has been measured in many biological and non-biological heterogeneous materials by means of the surface photovoltage effect, the vibrating probe technique, electrostatic force microscopy, among others [34,35]. Nevertheless, in the literature $\sigma_{12}$ at $\Sigma$ has not been experimentally measured nor calculated theoretically in cancer. Estimation of $\sigma_{12}$ at $\Sigma$ presupposes the experimental knowledge of normal components of the flux density vector on both sides of $\Sigma$, a procedure that is cumbersome and expensive (in time and resources) in preclinical and clinical studies. Furthermore, an analysis of TGK in terms of $\eta_k$, $\varepsilon_k$ (k = 1,2) and $\sigma_{12}$ has not been reported in the literature. $\sigma_{12}$ is ignored and cannot be estimated from the vast experimental data available. The aforementioned are the aspects we mainly take into consideration for using the physic-mathematical modeling in our research. Therefore, the aim of this study is to propose an approximate analytical expression that permits to simulate $\sigma_{12}$ at $\Sigma$ during the untreated tumor growth, in terms of two tumor kinetic parameters, tumor radius, and electrical properties of the tumor and its surrounding healthy tissue.

## 2    Methods

### 2.1    Assumptions

1. There is a three-dimensional, conductive, anisotropic and heterogeneous region consisting of two linear, anisotropic and heterogeneous media (tumor and the surrounding healthy tissue) separated by an interface $\Sigma$ (figure 1). Untreated solid tumor (medium inside $\Sigma$, named medium 1) is considered as a heterogeneous conducting sphere of radius $R_T$ (in m) of constant mean conductivity ($\eta_1$, in S/m) and mean permittivity ($\varepsilon_1$, in F/m). The surrounding healthy tissue (medium outside $\Sigma$, named medium 2) is supposed to be a heterogeneous infinite medium of constant mean conductivity ($\eta_2$, in S/m) and mean permittivity ($\varepsilon_2$, in F/m), where $\eta_1 > \eta_2$ and $\varepsilon_1 > \varepsilon_2$.

2. The source of electricity is neglected because the tumor is unperturbed.

3. Maxwell-Wagner-Sillars effect occurs physiologically between the tumor and the surrounding healthy tissue (see Introduction section).

4. In a first approximation, the electromotive force field ($\vec{E}_f$) depends only on the distance to the tumor center.

5. Normal and cancer cells that are at $\Sigma$ do not significantly contribute to $\vec{E}_f$.

### 2.2    Further comments about assumptions 1, 4 and 5

The first assumption may be approached from the physical point of view. As solid tumor and surrounding healthy tissue are anisotropic and heterogeneous media (formed by cells, water, ions, molecules, macromolecules, among others) [10,29,30], we consider that $\overline{\overline{\eta}}$ and $\overline{\overline{\varepsilon}}$ are real symmetric second-order tensors (3 x 3 matrix symmetric) of electrical conductivity and electrical permittivity, respectively. When the electrical conductivity and electrical permittivity are referred to the principal axes and both the electric field and the current density are related to the same coordinate system, then all nondiagonal elements are equal to zero and this 3 x 3 symmetric matrix becomes diagonal.

Consequently, there is an orthonormal base (which defines the so-called principal axes of the medium) in which $\overleftrightarrow{\eta}$ is represented by the diagonal matrix diag($\eta_1,\eta_2,\eta_3$), where $\eta_1$, $\eta_2$ and $\eta_3$ are electrical conductivities according to these main axes. If these diagonal elements are replaced by their mean value, named $\eta$ ($\eta = (\eta_1 + \eta_2 + \eta_3)/3$) in this approximation, the tensor $\overleftrightarrow{\eta}$ corresponds to the scalar matrix $\eta I$, where I is the identity matrix of order 3, as in [30]. The tensor $\overleftrightarrow{\varepsilon}$ is treated in the same manner and its mean value is $\varepsilon$.

Although cancer and surrounding healthy tissue are heterogeneous and anisotropic media, most experimental studies report their respective average values of $\overleftrightarrow{\eta}$ and $\overleftrightarrow{\varepsilon}$ tensors [6,10-14,28,29,36,37]. Furthermore, $\eta_1 > \eta_2$ and $\varepsilon_1 > \varepsilon_2$ have been explained because malignant tumors have a significantly higher water content, higher concentrations of ions and electrons, and altered cellular metabolism compared to those in the surrounding healthy tissue [6,8-10,15,16,19,38].

Tumor border plays a crucial role in growth, metastasis, aggressiveness, and anticancer therapy planning [1,4,38]. Locating them is not easy from a clinical point of view because the tumor border is a marginal zone that contains tumor cells and normal cells [39]. In this study, interface $\Sigma$ is the tumor-free margin, according to pathological anatomy reports. This ensures that there is no infiltration of the tumor into the surrounding healthy tissue and there are two well-defined regions instead. In addition, the geometry and border of the tumor (regular or irregular) have no relevance for this tumor-free margin; therefore, the tumor contour may be assumed regular, sharp and smooth. It should be noted that $\Sigma$ is not chosen as the surgical margin because it does not guarantee that the tumor has infiltrated adjacent normal tissue [39].

The term "infinite healthy tissue" does not mean an unlimited space but it rather refers to an enormous healthy tissue (the limited region free of infiltration and metastasis of cancer cells) in comparison to the tumor.

A standard pattern of three-dimensional anatomically realistic models (numerically solved with COMSOL-Multiphysics and similar packages) is very unlikely to be established in simulations because it requires a precise knowledge of the electric properties (i.e., electrical conductivity, electrical permittivity) and physiologic characteristics (i.e., type, heterogeneity, anisotropy, size, shape, composition, structure, consistency, and water content) of both tissues. This becomes even more cumbersome when other characteristics of the tumor are taken into account, such as: histological variety, stage, stiffness, mitotic index, degree of anaplasia, invasiveness and metastasis. In addition, our vast experience in preclinical studies evidence differences in space-time patterns of a same tumor histological variety that grows in several BALB/c/Cenp mice under the same experimental conditions (temperature and relative humidity of the room; initial concentration and viability of cancer cells; and mice with the same age, gender and weight) [1,2,4]. This result is due to the biological individuality. If this analysis is individualized, an individual model should be suggested for each patient/animal, which is not feasible from a theoretical and experimental standpoint. Furthermore, the diversity and complexity of non-spherical geometries [1-4] and irregular borders [7,10,38,40] of tumors during their growths make very difficult to establish a single spatiotemporal pattern of these two aspects for simulations. The electrical and biological parameters of the tumors cannot be controlled by the performing physician (in clinics) or researchers (*in vitro* and *in vivo* studies). That is why, tissue realistic conditions are not considered in this study.

The tumor spherical geometry is observed in 3D culture [41,42] and first time instants of TGK (tumor sizes $\leq 50$ mm$^3$) [1,2,4]. Spherical cancer models (major 3D *in vitro* models) have been used in cancer research as an intermediate model between *in vitro* cancer cell line cultures and *in vivo* tumor. These models have gained popularity in screening environments for better assessment and characterization of anticancer therapy efficacy (i.e., chemoresistance, radioresistance), identify potential cancer therapeutics, among others applications. Chemoresistance and radioresistance of cancer may be more marked in spherical tumors than those in non-spherical tumors, according to the results of the simulations reported by Castañeda et al. [43] and the sphere is the only geometry that is in contact with another surface at a point. Furthermore, they can be used as reliable models of *in vivo* solid tumors and drug screening platforms. Tumor spheroids may contribute to decrease animal experimentation [41,42]. The aspects and the poor understanding of $\sigma_{12}$ at $\Sigma$ from both experimental and theoretical points of view (unknown effects of the irregular border and changes in $\sigma_{12}$ at $\Sigma$) are the reasons why we use the tumor spherical shape and symmetry to know $\sigma_{12}$ at $\Sigma$ approximately.

Until now, space-time distributions of $\vec{E}_f$ are neither experimentally nor theoretically known. In this study, $\vec{E}_f$ represents the active bioelectricity of unperturbed cancer and due to endogenous electrical biopotentials ($\phi$) and/or intrinsic electrical sources in it [8, 9,14-16,19-26]. Miklavčič et al. [44] measure experimentally $\phi$ along axial (z-axis) and radial (r-axis) directions in two tumor types (LLC and fibrosarcoma Sa-1). They report several findings, such as: $\phi$ is negative in entire tumor; $\phi$ is more electronegative in the tumor center (-160 mV for LLC tumor and -131.5 mV for fibrosarcoma Sa-1); electronegative of $\phi$ is less negative towards the periphery in LLC and fibrosarcoma Sa-1 tumors; and values of $\phi$ depend on distance (from tumor center to its periphery) and not on angular coordinates. It is important to point out that $\phi$ should not be confused with the electric potential applied to a tissue by means of electrodes [2,30]. These are the reasons why the above mentioned fourth assumption is proposed.

Although tumor border is a marginal zone that contains tumor cells and normal cells from a clinical point of view, cancer cells at $\Sigma$ invade the surrounding healthy tissue and do not migrate to the tumor interior. As a result, these cells do not contribute to $\vec{E}_f$. As $\vec{E}_f$ is only related to the unperturbed cancer active bioelectricity, normal cells at $\Sigma$ do not contribute to $\vec{E}_f$. In addition, normal cells at $\Sigma$ are transformed in cancer cells during tumor growth. These aspects justify why the fifth assumption.

## 2.3 Theory

The assumptions in section 2.1 and the close relationship between physical and biological aspects in cancer allow us to consider that $\phi$ and $\vec{E}_f$ are related, in a first approximation, by means of the equation

$$\nabla \bullet \vec{\eta} \bullet \left(-\nabla\phi + \vec{E}_f\right) = 0, \tag{1}$$

where $\vec{\eta}$ is the symmetric second-order tensor of the electrical conductivity of any linear, anisotropic and non-homogeneous medium (for example, a biological tissue). This tensor is used in previous studies [29,30].

Equation (1) is obtained by combining the continuity equation ($\nabla \bullet \vec{J} + \partial\rho/\partial t = 0$) for the static case ($\partial\rho/\partial t = 0$) and law of Ohm ($\vec{J} = \vec{\eta} \bullet (\vec{E} + \vec{E}_f)$), valid for media of linear conduction. In this case, $\vec{J} = \vec{J}(\vec{r})$ is the electric current density $\vec{J}(\vec{r}) = \rho(\vec{r})\vec{v}(\vec{r})$, where $\rho(\vec{r})$ is the electric charge density and $\vec{v}(\vec{r})$ the velocity field of electric current carriers.

Assumptions 2-5 in section 2.1 allow considering that $\vec{D} = \varepsilon\vec{E}$ and the medium is considered isotropic in this approach, where $\vec{D}$ is the induction field (flux density vector). Taking this into account, and assuming that the medium is electrically homogeneous, equation (1) has the form

$$\nabla \bullet \eta\left(-\nabla\phi + \vec{E}_f\right) = \eta\nabla \bullet \left(-\nabla\phi + \vec{E}_f\right) = \eta\left[-\nabla^2\phi + \nabla \bullet \vec{E}_f\right] = 0. \tag{2}$$

Therefore,

$$\nabla^2\phi = \nabla \bullet \vec{E}_f. \tag{3}$$

## 2.4 Boundary conditions

# Surface charge density changes in tumor growth

The region of interest is assumed as a heterogeneous biological tissue formed by the solid tumor (with average values $\eta_1$ and $\varepsilon_1$) surrounded by the surrounding healthy tissue (with average values $\eta_2$ and $\varepsilon_2$), as shown in figure 1.

According to the continuity equation for the static case, the current density normal components of the tumor ($J_{1n}$) and the surrounding healthy tissue ($J_{2n}$) are continuous at $\Sigma$

$$J_{1n} = J_{2n}. \tag{4}$$

Therefore,

$$\eta_1 E_{1n} = \eta_2 E_{2n} \Longrightarrow \eta_1 \frac{\partial \phi_1}{\partial n} = \eta_2 \frac{\partial \phi_2}{\partial n}, \tag{5}$$

where $E_{1n}$ is the normal component of the electrical field of the tumor. $E_{2n}$ is the normal component of the electrical field of the surrounding healthy tissue. $\phi_1$ is the electrical potential in the tumor and $\phi_2$ the electrical potential in the surrounding healthy tissue. The normal derivatives of $\phi_1$ and $\phi_2$ are $\partial \phi_1 / \partial n$ and $\partial \phi_2 / \partial n$, respectively.

Equation (4) is valid if $\vec{E}_f = \vec{0}$ at $\Sigma$ (see Assumption 9). The positive normal to the tumor surface is indicated as a unit vector $\vec{n}$ (represented schematically in figure 1 by **n**) draw from the surrounding healthy tissue (medium 2) into the tumor (medium 1). According to this convention, medium 2 lay on the negative side ($\vec{n}_2 = -\vec{n}$), and medium 1 on the positive side ($\vec{n}_1 = \vec{n}$). Taking this into account as well as the matching boundary condition for $\vec{D}$, $D_{1n} - D_{2n} = \sigma_{12}$, equation (5) and $\vec{D} = \varepsilon\vec{E}$ result for $\sigma_{12}$ the expression

$$\sigma_{12} = \varepsilon_2 \left[ \frac{\eta_1}{\eta_2} - \frac{\varepsilon_1}{\varepsilon_2} \right] \frac{\partial \phi_1}{\partial n} = -\varepsilon_2 \left[ \frac{\eta_1}{\eta_2} - \frac{\varepsilon_1}{\varepsilon_2} \right] E_{1n}, \tag{6}$$

where $D_{1n}$ and $D_{2n}$ are the normal components of the flux density vector $\vec{D}$ in the tumor and the surrounding healthy tissue, respectively.

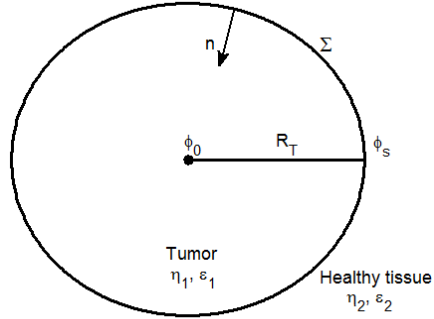

**Figure 1. Schematic representation of a spherical tumor surrounded of its healthy tissue.** Variables $\phi_0$ and $\phi_s$ denote the electrical potentials in the center and the periphery of the tumor, respectively. $R_{T0}$ is the initial tumor radius. $\eta_i$ and $\varepsilon_i$ represent the electrical conductivities and electrical permittivities of the tumor ($i = 1$) and the surrounding healthy tissue ($i = 2$). **n** denotes the inward unit normal vector to the boundary $\Sigma$ (interface that delimits both tissues).

## 2.5 Calculation of the free electric charge surface density $\sigma_{12}$

Strictly speaking, the problem to be solved for the calculation of the electric potential is equation (3) subject to the matching boundary conditions for $\phi$ and $\partial \phi / \partial n$

$$\begin{cases} \nabla^2 \phi = \nabla \bullet \vec{E}_f(\vec{r}) \\ \phi_1 = \phi_2 \\ \eta_1 \frac{\partial \phi_1}{\partial n} = \eta_2 \frac{\partial \phi_2}{\partial n} \end{cases}, \tag{7}$$

where $\vec{r} \in \Sigma$.

The tumor spherical model is reported in [41,42]. As $\phi$ at $\Sigma$ may be experimentally measured, conditions that may be replaced by a condition of Dirichlet and the work region is only inside the spherical tumor, of radius R, the solution of the problem of Poisson into the tumor in spherical coordinates (r,θ,φ; $0 \leq r < R$, $0 \leq \theta \leq \pi$ and $0 \leq \varphi \leq 2\pi$) is given by

$$\phi_1(r,\theta,\varphi) = \phi_{1h} + \phi_{1p} = \sum_{n=0}^{\infty}\sum_{m=0}^{n} r^n P_n^m(\cos\theta)(A_{nm}\cos m\varphi + B_{nm}\operatorname{sen} m\varphi) + \phi_{1p}, \qquad (8)$$

where $P_n^m(\cos\theta)$ are the generalized polynomials of Legendre and $\phi_{1p}$ is a particular solution any of equation (3) in the tumor.

Assumption 7 supposes that $\nabla \bullet \vec{E}_f = 2/r$. In this case, the solution (8) is bounded and it does not depend on the coordinates θ and φ, given by

$$\phi_1 = A\frac{r}{R} + A_{00}. \qquad (9)$$

Constants A and $A_{00}$ in equation (9) are calculated from $\phi_0 = \phi_1(r=0)$ and $\phi_s = \phi_1(R,0,0)$, being $A = \phi_s - \phi_0$ and $A_{00} = \phi_0$. As a result, $\phi_1(r)$ is given by

$$\phi_1(r) = \frac{\phi_s - \phi_0}{R}r + \phi_0, \quad 0 \leq r \leq R \qquad (10)$$

In equation (10), the difference between $\phi_0$ and $\phi_s$ represents the tumor heterogeneity from the electrical point of view. The term $(\phi_0 - \phi_s)/R$ is interpreted as the linear radial gradient of $\phi_1(r)$.

The electric field intensity in the tumor is calculated from $\vec{E}_1 = -\nabla\phi_1$, given by

$$E_1(r) = \frac{\phi_0 - \phi_s}{R}, \quad 0 \leq r \leq R. \qquad (11)$$

Equation (11) shows that the electric field is uniformly distributed in the entire tumor volume. If equation (11) is substituted in equation (6), the following expression is found for $\sigma_{12}$, given by

$$\sigma_{12} = -\varepsilon_2\left[\frac{\eta_1}{\eta_2} - \frac{\varepsilon_1}{\varepsilon_2}\right]\left[\frac{\phi_0 - \phi_s}{R}\right]. \qquad (12)$$

Equation (12) gives the dependence of $\sigma_{12}$ with $\phi_0$, $\phi_s$, R, $\eta_k$ and $\varepsilon_k$ (k = 1,2) for a fixed time after tumor cells are inoculated in the organism. R is any tumor radius higher and equal than $R_m$, where $R_m$ is the minimum tumor radius measured in preclinical studies or the first tumor radius detected in clinics [1,2]. The term $(\eta_1/\eta_2 - \varepsilon_1/\varepsilon_2)$ represents the difference between the conductive and dielectric ratios of the tumor and the surrounding healthy tissue.

Several experimental studies report that R of untreated tumors changes in the time t [1-3]. As a result, $\sigma_{12}$ is expected to depend on t. For this, CGE is used.

## 2.6    Conventional Gompertz equation

CGE is given by

$$V_T(t) = V_{T0}e^{\left(\frac{\alpha}{\beta}\right)(1-e^{-\beta t})}, \qquad (13)$$

where $V_T(t)$ represents the tumor volume at a time t after tumor cells are inoculated into the host. The initial tumor volume ($V_{T0}$) is given by the initial condition $V(t=0) = V_{T0}$. The parameter α (α > 0) is the intrinsic growth rate of the tumor. The parameter β (β > 0) is the growth deceleration factor due to endogenous antiangiogenic process [1,2,4].

As the tumor is assumed a spheroid, $V_T(t)$ in CGE corresponds to the volume of a sphere ($V_T(t) = 4\pi R_T^3(t)/3$, where $R_T(t)$ is the spheroid tumor radius at a time t). As $R_T(t)$ and $V_T(t)$ depend on t, R in equation (12) is replaced by $R_T(t)$. As a result, $\sigma_{12}$ is a function of t, named $\sigma_{12}(t)$. Substituting $V_T(t)$ in equation (13) results

$$R_T(t) = R_T = R_{T0}\sqrt[3]{e^{\left(\frac{\alpha}{\beta}\right)(1-e^{-\beta t})}}, \qquad (14)$$

where $R_{T0}$ satisfies the initial condition $R_T(t=0) = R_{T0}$ (figure 1).

The substitution of equation (14) in equation (12) allows to express approximately $\sigma_{12}$ in terms of $R_{T0}$, $\phi_0$, $\phi_s$, $\eta_1$, $\varepsilon_1$, $\eta_2$, $\varepsilon_2$, i, $i_0$, $\alpha$, $\beta$ and t, unprecedented in the literature. In this study, three graphs for $R_T$ ($R_T$ versus t, $dR_T/dt$ versus t, and $dR_T/dt$ versus $R_T$) and three graphs for $\sigma_{12}$ ($\sigma_{12}$ versus $R_T$, $d\sigma_{12}/dt$ versus t, and $d\sigma_{12}/dt$ versus $\sigma_{12}$) are analyzed, where $dR_T/dt$ is the first derivative of $R_T$ with regard to t whereas $d\sigma_{12}/dt$ is the first derivative of $\sigma_{12}$ with respect to t. From these six graphs, four graphs are only shown in this study: $R_T$ versus t, $dR_T/dt$ versus $R_T$, $\sigma_{12}$ versus $R_T$, and $d\sigma_{12}/dt$ versus $\sigma_{12}$.

## 2.7    Simulations

For simulations, we use values of $\alpha$ (0.6 days$^{-1}$) and $\beta$ (0.2 days$^{-1}$) corresponding to the fibrosarcoma Sa-37 tumor [1,2], $R_{T0}$ (5.6 mm) and $\phi_0$ = -160 mV corresponding to LLC tumor [44], and different values of $\phi_s$ (between -15 and -135 mV) and $\eta_1/\eta_2$ - $\varepsilon_1/\varepsilon_2$ (between 1 and 5). For these values of $\phi_0$ and $\phi_s$, $\phi_0$ - $\phi_s$ varies between -145 and -25 mV. In this study, we only show results for $\phi_0$ - $\phi_s$ (-145 and -25 mV) and $\eta_1/\eta_2$ - $\varepsilon_1/\varepsilon_2$ (1, 3 and 5). Furthermore, the parameter $\varepsilon_2$ in equation (12) is calculated by the expression $\varepsilon_2 = \varepsilon_{r2}\varepsilon_0$, where $\varepsilon_0$ (8.85x10$^{-12}$ F/m) is the vacuum permittivity and $\varepsilon_{r2}$ (4x10$^7$) the relative permittivity of the muscle. Muscle is one of tissues where tumor cells are more frequently inoculated subcutaneously [1,2]. This is why, the muscle and its electrical properties are chosen in this study to characterize the healthy tissue that surrounds the tumor.

The aforementioned range of $\phi_s$ aforementioned may be justified for the following three reasons. First, $\phi_s$ is unknown experimentally and theoretically. Second, $\phi$ are less negative towards the peripheries of LLC and fibrosarcoma Sa-1 tumors [44]. Third, approximate knowledge of how $\sigma_{12}$ at $\Sigma$ is affected by difference of $\phi$ between the center and border of tumor from bioelectrical point of view. This is taken into account because the strongest tumor aggressiveness shows its higher difference between the center and border of the tumor from an oncological point of view [39]. That is why, we do not use $\phi_0$ = -131.5 mV (for fibrosarcoma Sa-1 tumor) [44] for simulation. Fourth, $\phi_s$ depends on the histological variety and size of the tumor, organ/tissue where it grows, type of medium (cell culture, ex vivo tissue, or organism (i.e., animal, body human)).

Many authors report $\eta_1$, $\eta_2$, $\varepsilon_1$ and $\varepsilon_2$ values for different tumor histological varieties [6,10-12,28,36]. We calculate $\eta_1/\eta_2$ and $\varepsilon_1/\varepsilon_2$ ratios for each tumor type and all satisfy that $0 < \eta_1/\eta_2$ - $\varepsilon_1/\varepsilon_2 < 5$. $\eta_1/\eta_2$ - $\varepsilon_1/\varepsilon_2$ =0 ($\eta_1\varepsilon_2 = \varepsilon_1\eta_2$) supposes that the tumor and surrounding healthy tissue have the same electrical properties, in contrast with the experiment [6,10-12,28,36]. If $\eta_1/\eta_2$ - $\varepsilon_1/\varepsilon_2$ increases, the conductor properties prevail in both tissues; therefore, they behave as electrical conductors, being marked for the tumor. Contrastively, the conductor and dielectric properties prevail in these two tissues when $\eta_1/\eta_2$ - $\varepsilon_1/\varepsilon_2$ is small. In this case, both tissues behave as real dielectrics. This may be relevant in the aggressiveness and therapeutic planning of tumors [29,30]. These are the reasons why we varied $\eta_1/\eta_2$ - $\varepsilon_1/\varepsilon_2$ between 1 and 5.

A computer program is implemented in the Matlab® software (version R2012b 64-bit, University Institute for Research in Mathematics and Applications, University of Zaragoza, Zaragoza, Spain) to calculate and simulate the tumor radius, free electric charge surface density and their first derivate in time. These calculations are performed on a PC with an Intel(R) core processor (TM) i7-3770 at 3.40 GHz with a Windows 10 operating system. All calculations take approximately 1 min.

## 3    Results

Figure 2 shows simulations of $R_T$ versus t (figure 2a) and $dR_T/dt$ versus $R_T$ (figure 2b). Nevertheless, figure 3 reveals simulations of $\sigma_{12}$ versus $R_T$ (figure 3a,b) and $d\sigma_{12}/dt$ versus $\sigma_{12}$ (figure 3c,d). The simulations of $\sigma_{12}$ versus $R_T$ and $d\sigma_{12}/dt$ versus $\sigma_{12}$ are shown for three values of $\eta_1/\eta_2 - \varepsilon_1/\varepsilon_2$ above-mentioned and two values of $\phi_0 - \phi_s = -145$ mV (figure 3$a$,$c$) and $-25$ mV (figure 3$b$,$d$).

The simulations of $R_T$ versus t and $\sigma_{12}$ versus t have similar behaviors (figure is not shown). When time elapsed, $R_T$ and $\sigma_{12}$ grow up to their asymptotic values reached for t = 40 days, called $R_{T-f}$ and $\sigma_{12-f}$, respectively. The value of $\sigma_{12-f}$ (stationary condition for $\sigma_{12}$) is less negative than $\sigma_{12-0}$ and its value depends on $\phi_0 - \phi_s$ and $\eta_1/\eta_2 - \varepsilon_1/\varepsilon_2$, where $\sigma_{12-0}$ is the value of $\sigma_{12}$ at t = 0. Although the graphs of $dR_T/dt$ versus t and $d\sigma_{12}/dt$ versus t are not shown in this study, it can be proved that both graphs evidence similar behaviors. These graphics show that positive values of $dR_T/dt$ and $d\sigma_{12}/dt$ decrease asymptotically to zero when time increases. Figure 2$b$ reveals that $dR_T/dt$ decreases non-linearly to zero when $R_T$ increases, while $d\sigma_{12}/dt$ decreases when $\sigma_{12}$ is less negative (figure 3$c$,$d$). In addition, two stages are identified from the graphic strategies shown in Figures 2 and 3: the first grows rapidly (positive slope) and the second stationary ($R_T$ and $\sigma_{12}$ are constant over time).

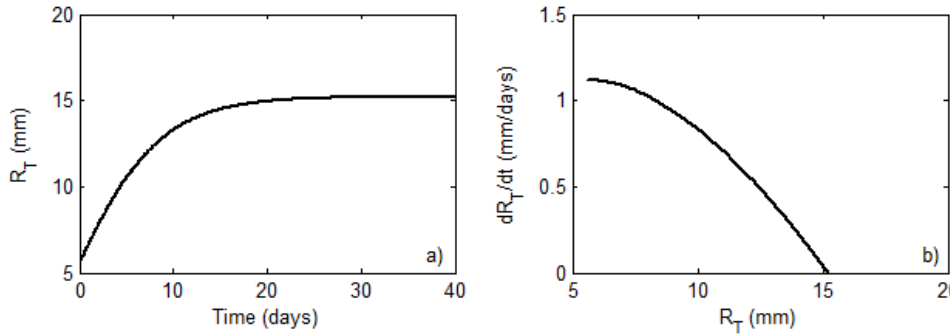

**Figure 2. Unperturbed tumor radius.** Simulations of (a) $R_T$ against time t, (b) $dR_T/dt$ versus $R_T$.

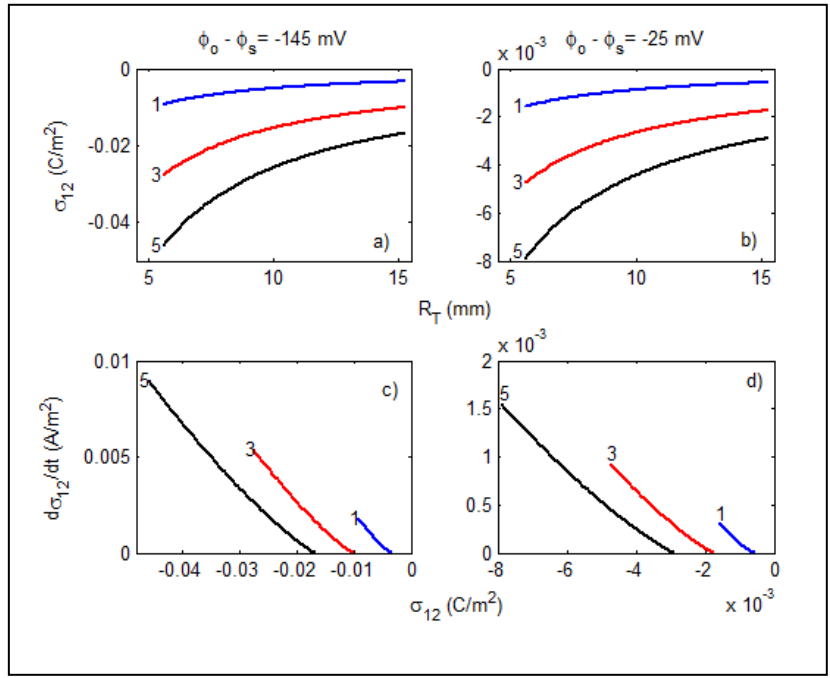

**Figure 3. Free electric charge surface density in unperturbed tumor.** Simulations of (a) $\sigma_{12}$ versus $R_T$ for $\phi_0 - \phi_s = -145$ mV, (b) $\sigma_{12}$ versus $R_T$ for $\phi_0 - \phi_s = -25$ mV, (c) $d\sigma_{12}/dt$ versus $\sigma_{12}$ for $\phi_0$ - $\phi_s$ = -145 mV, (d) $d\sigma_{12}/dt$ versus $\sigma_{12}$ for $\phi_0$ - $\phi_s$ = -25 mV. Three values of $\eta_1/\eta_2$ - $\varepsilon_1/\varepsilon_2$ (1, 3 and 5) are shown in each sub-plot.

## 4.    Discussion

Although endogenous electric potentials and electrical properties of the cancer and surrounding healthy tissue may be measured [10-14,28], $\sigma_{12}$ at $\Sigma$ has not been experimentally measured or theoretically calculated for cancer. That is why, our simulations have not been experimentally validated (main limitation of this study) nor stochastically. Although stochastic simulation models are used to describe TGK [45,46], they are black box and complex. Furthermore, random variations in stochastic models (due either to uncertainties on the parameter or to small population sizes) may influence on value of $\sigma_{12}$ at $\Sigma$, but do not change its time behavior. These aspects have made us use deterministic models in this study. And these models are feasible to describe TGK [1-4], they are also simple, easily understandable, and more appropriate for some customers. They also comprise a known set of inputs (i.e. $\alpha$, $\beta$, $R_{T0}$, $\phi_0$, $\phi_s$, $\eta_1$, $\eta_2$, $\varepsilon_1$, and $\varepsilon_2$ (or $\varepsilon_0$ and $\varepsilon_{r2}$)) which will result in an unique set of outputs (i.e., $R_T$ and $\sigma_{12}$). In our approach, all random variations are implicitly included in parameters of $\sigma_{12}$ (i.e., $\alpha$, $\beta$, $\eta_1$, $\eta_2$, $\varepsilon_1$ and $\varepsilon_2$).

If "realistic simulations" are taken into account [45,46], the simple mathematical approach, "nonrealistic tissue" and tumor non-spherical geometry used in this study may represent a restraint for many researchers; nevertheless, we must be careful with this statement (see our comments in subsection 2.2). The assumptions in subsection 2.1 are reasonable and supported by experimental studies. Furthermore, the results of our formalism agree (in good approximation) with experimental and theoretical results reported in the literature (see below) and suggest other findings not considered recently. Therefore, the results of our biophysic-mathematical approach are valid for such considerations. The novelty of this study does not lie in the use of Ohm law, Poisson and conventional Gompertz equations and problem of boundary conditions between two dielectric media as these are well-known facts.

This study has two main achievements. First of all, the simple biophysic-mathematical approach proposed in this study that allows to know an approximated theoretical expression that relate $\sigma_{12}$ at $\Sigma$ with tumor parameters ($V_{T0}$, $\alpha$ and $\beta$), tumor electrical properties ($\eta_1$ and $\varepsilon_1$), bioelectrical potential in the tumor ($\phi_0$, $\phi_s$), electrical properties of surrounding healthy tissue ($\eta_2$ and $\varepsilon_2$), which is unprecedented in the literature. Likewise, explicit knowledge of $\sigma_{12}$ at $\Sigma$ with $\alpha$ and $\beta$ allows to relate $\sigma_{12}$ at $\Sigma$ in terms of Avrami exponent and impingement parameter [1]; apoptosis rate, fractal dimension of the tumor contour, and fractal dimension of tumor mass [4,5], unprecedented in the literature too. Furthermore, this approach constitutes a rapid and simple method for visualizing both $R_T$ and $\sigma_{12}$ at $\Sigma$ changes in time without using special software for numerical modeling. That is another reason why we prefer the analytical method. Second, researchers in cancer should take into account our results to increase the effectiveness of anticancer therapies, mainly chemotherapy, immunotherapy and physical therapies (i.e., electrochemical therapy, electroporation irreversible, hyperthermia, electrochemotherapy).

The results of this study confirm several findings reported in the literature and suggest others not yet revealed, such as: $\sigma_{12} \neq 0$ at $\Sigma$ is a direct consequence of equation (12) if ($\eta_1\varepsilon_2$ - $\eta_2\varepsilon_1$) $\neq 0$ and corroborates the existence of a multi-system with two different loss dielectrics in contact: the solid tumor and surrounding healthy tissue. Loss dielectric is a dielectric that has finite electrical conductivity and its induced electrical charges can move but not as freely as they would in a perfect conductor. If $\sigma_{12} = 0$ at $\Sigma$ ($\eta_1\varepsilon_2$ - $\eta_2\varepsilon_1 = 0$), $\eta_1 = \eta_2$ and $\varepsilon_1 = \varepsilon_2$, in contrast with the experiment [10,13,14,28-30]. Therefore, $\sigma_{12}$ at $\Sigma$ must be considered in cancer and its TGK.

# Surface charge density changes in tumor growth

As $\sigma_{12} \neq 0$ at $\Sigma$, there are average current densities $J_1$ (in the entire tumor volume V ($\int_V \vec{J}_1 dV \neq 0$) and a consequence of $\nabla \bullet \vec{E}_f \neq 0$) and $J_2$ (in the surrounding healthy tissue), being $|J_1| > |J_2|$ because $\eta_1 > \eta_2$ [10,11,14,36]. This corroborates that electrical properties, active bioelectricity (i.e., $\phi$, concentrations and mobility of electrical charges) are much higher in the tumor than those in the surrounding healthy tissue [8,9,11,15]. $J_1$ and $J_2$ on both sides of $\Sigma$ indicate the Maxwell-Wagner-Sillars effect occurs for the tumor-surrounding healthy tissue multi-system. Due to this effect, both free and bound surface charge densities contribute to $\sigma_{12}$ (interfacial polarization). Therefore, the Maxwell-Wagner-Sillars effect must not be ignored in cancer. The motion of electrical charges in both biological tissues involved during the tumor growth happen in different time scales, named relaxation times ($\tau$), being $\tau_1$ for the tumor ($\tau_1 = \varepsilon_1/\eta_1$ and it depends on the tumor histological variety) and $\tau_2$ for the surrounding healthy tissue ($\tau_2 = \varepsilon_2/\eta_2$ and it depends on the tissue type). These aspects may suggest that both tissues cannot be perfect conductors ($\tau_1$ and $\tau_2$ tend to zero because $\eta_1$ and $\eta_2$ are infinite) or perfect dielectrics (induced volume charges cannot move), corroborating that these two tissues are loss dielectrics.

$\nabla \bullet \vec{E}_f \neq 0$ considers that the tumor heterogeneity is implicit in the model, but not for the intra-tumor anisotropy. If the tumor and surrounding healthy tissue are assumed anisotropic, electrical properties of these two tissues should be replaced by their corresponding tensors. As a result, equations must be replaced by more complicated ones, being cumbersome the calculation procedure for obtaining the analytical solution of the problem (8). As $\nabla \bullet \vec{E}_f$ is positive through the tumor interior, negatively charged electrical sources prevail ($J_1 < 0$), corroborating the tumor electronegativity (negative electric bio-potentials) [12-14] and the ionic and faradic currents should not be analyzed separately [8]. This may indicate that positively charged carriers may be directed from the tumor towards the surrounding healthy tissue, explaining its electropositivity ($J_2 > 0$) [13,14].

It should not be ignored that $J_1$ may create a macroscopic magnetic field into the entire tumor and therefore an endogenous magnetic energy (per unit volume) that grows rapidly with increasing tumor size, in agreement with electric and magnetic fields (static or variable in time) associated with constant and time-varying endogenous electrical currents [8-10]. All these physical magnitudes are weak due to the breakdown of intercellular communication in the cancer [11,14,18] and theoretically corroborated here because $\nabla \bullet \vec{E}_f = 2/r$ (divergence of $\vec{E}_f$ decreases when $r \to R_T$). This corroborates that $\vec{E}_f$ is weak and a weak electrical coupling between cancer cells, mainly into tumor regions near $\Sigma$, due to the higher electrophysiological activity of them is in these regions, as documented in [1,4,5,7,13,14,18,33]. Migration of positively charged carriers from the tumor to the surrounding healthy tissue makes that ionic bridges (strong interaction) among negatively and positively charged carriers are not formed and therefore weak interactions among cancer cells. Weak signals from biological systems are reported in [47]. If $\vec{E}_f = 0$, the tumor dies. $\nabla \bullet \vec{E}_f = 2/r$ indicates that the highest electronegativity is in the tumor center because $\vec{E}_f$ is very intense in $r = 0$, aspect that may explain in part the endogenous central intra-tumor necrosis and migration of tumor cells towards $\Sigma$. For this, $\vec{E}_f$ should be higher and equal than the endogenous physiological electric field in tumors. Central intra-tumor necrosis explained here from the electrical point of view does not contradict explanations given to it related to the lack of oxygen and nutrients in the central region of the tumor during its growth [8,48].

The time variation of $\sigma_{12}$ at $\Sigma$ corresponds to the change from the quick tumor growth phase to asymptotic phase of TGK and follows a sigmoidal behavior in time, as TGK [1-5]. This non-linear time behavior of $\sigma_{12}$ may be explained because $\eta_1$ and $\eta_2$ exhibit non-linear behavior due to biological tissues are non-linear systems [1,4], and $\eta_k$, $\varepsilon_k$ (k = 1,2), $\tau_1$, $\tau_2$ and the relaxation time of the interfacial polarization ($\tau_p$) change in time [49]. Furthermore, these physical magnitudes, $\phi_o$, $\phi_s$ and ($\phi_o$ - $\phi_s$) may change in time by dynamic self-regulation of $\sigma_{12}$ at $\Sigma$; nevertheless, we do not explicitly know how? Therefore, we assume constants these physical parameters in our approximation, being a limitation of our model.

This dynamic change of $\sigma_{12}$ at $\Sigma$ must be self-regulated during tumor growth, as TGK [1,4]. It is faster for the most undifferentiated tumors (most aggressive: greater difference of $\alpha$ with respect to $\beta$), strong endogenous electrical potential gradient in cancer (greater permissible difference between $\phi_o$ and $\phi_s$) and the greater difference is between ratios of electrical properties of the tumor and surrounding healthy tissue (maximum permissible value of $\eta_1/\eta_2$ - $\varepsilon_1/\varepsilon_2$). This endogenous electrical potential gradient may explain the altered cancer bioelectricity (i.e., higher mobility of ions, electrons, charged molecules and cancer cells) [8,19]. The change from $\sigma_{12-0}$ (more negative) to $\sigma_{12-f}$ (less negative) at $\Sigma$ suppose that negativity of $\sigma_{12}$ changes dynamically over time during TGK, being marked for the highest values of $\phi_0$ - $\phi_s$ and $\eta_1/\eta_2$ - $\varepsilon_1/\varepsilon_2$. This finding may impact in both chemical and electrical environments of the cancer cells and the solid tumor; the hypocellular gap on the tumor-host interface (responsible of the differentiation between tumor electrical properties and the surrounding healthy tissue) [50]; $\tau_p$, which depends on $\eta_k$ and $\varepsilon_k$ (k = 1,2) [31,32]; and spatiotemporal dynamic at $\Sigma$ [51].

The tumor electronegativity during its growth ($\nabla \bullet \vec{E}_f > 0$) may be explained from generation of more negative charges produced by different redox processes, duplication of cancer cells (mainly in regions near $\Sigma$) and/or the dynamic self-regulation of $\sigma_{12}$ at $\Sigma$ (molecules and ions negatively charged, and electrons migrate in time from $\Sigma$ towards the entire tumor interior until $\sigma_{12}$ = $\sigma_{12-f}$). This may suggest that dynamical alterations in cancer bioelectricity impact in its growth, invasion, metastasis, maximum survival, neutralization of the attack of the immune system, and resistance to anticancer therapies, as report in previous studies [8,12-15,33,52-55]. This dynamic self-regulation of $\sigma_{12}$ at $\Sigma$, $\nabla \bullet \vec{E}_f = 2/r$, and cancer cells negatively charged [14] may suggest three aspects. First, the electrostatic repulsion among cancer cells facilitates migration, invasion and metastasis of them [53]. For this, electric biopotentials have to be more negative in the central region of the tumor than in its periphery, during its growth over time, as in [44], so that the entire tumor behaves like a negatively charged heterogeneous endogenous electrical shield. Second, electric field intensity of this shield changes dynamically over time. It depends on $\phi_0$ - $\phi_s$, $\eta_1/\eta_2$ - $\varepsilon_1/\varepsilon_2$ and dynamic change in $\sigma_{12}$ at $\Sigma$, and electrostatically repels humoral and cellular components of the immune system, mainly those negatively charged (e.g., T lymphocytes, natural killer cells, among others) [14]. Consequently, the immune system does not recognize the tumor. Third, positive electrical charges migrate toward the surrounding healthy tissue, as reported for diffusion of hydrogen ions, which damage the normal tissue [50]. This migration of positively charged carriers through $\Sigma$ may avoid that $\sigma_{12}$ = 0 at $\Sigma$ and weaken the electrostatic coupling among cancer cells (negatively charged) in tumor regions near $\Sigma$ to favor metastasis of them. This may explain the acidification of the tumor microenvironment, which is related to the progression, invasion, metastasis, stimulation of many immunosuppressive processes, and resistance to anticancer therapies [8].

If $\sigma_{12-f}$ were much more negative than $\sigma_{12-0}$ at $\Sigma$, the tumor would behave as an isolated system because carriers of negative electrical charges would essentially concentrate at $\Sigma$. This interface behaves as an electrical barrier that prevents the entry and exit of different substances through it. If $\sigma_{12}$ = 0 at $\Sigma$, the cellular elements of the immune system would enter the tumor interior. In both cases, the tumor would completely self-destruct, in contrast with the experiment [1,4,5]. Endogenous angiogenesis may be the emerging physiological mechanism to avoid $\sigma_{12}$ = 0 at $\Sigma$ and replace the mechanism for which $\sigma_{12}$ at $\Sigma$ changes from $\sigma_{12-0}$ to $\sigma_{12-f}$ as the tumor increases in size. This latter facilitates the migration of cancer cells towards the surrounding healthy tissue and the entry of nutrients into the tumor during its growth because blood is the most conductive tissue in the human body. This may justify why angiogenesis process in cancer emerges due to changes in its electrical and mechanical parameters at $\Sigma$, as previously reported in [1,4].

Although malignant tumors are not generally spherical [1-4,7], results of this study confirm the usefulness of the spheroidal model of a tumor to reveal intrinsic findings in its TGK, in accordance with [11,49]. If we consider that boundary condition depends on the spherical coordinates $(r,\theta,\varphi)$ in problem (8), $\sigma_{12}$ would depend on $(R_T,\theta,\varphi)$, which means that $\sigma_{12}$ is not uniform at the entire $\Sigma$. Furthermore, $\phi_1(r)$, $E_1(r)$ and $J_1(r)$ depend nonlinearly on $(r,\theta,\varphi)$. As the ellipsoidal geometry of the solid tumor is often observed in the experiment [1-4,7], the problem (8) has to be solved in elliptical coordinates. For a tumor arbitrary geometry, solution of the problem (8) is more complex and requires numerical methods.

The results of this study evidence that $R_T$ and $\sigma_{12}$ at $\Sigma$ change in time during tumor growth for constant values of $\alpha$, $\beta$, $\phi_0$, $\phi_s$, $\eta_1$, $\eta_2$, $\varepsilon_1$, and $\varepsilon_2$. These eight parameters as well as $\phi_1(r)$, $E_1(r)$, and $J_1(r)$ it are expected to change in time too due to biological changes in tumor growth, as necrosis (central or no), angiogenesis, among others. Nevertheless, there are no relevant experimental/theoretical information available that link these two biological findings with $\sigma_{12}$ at $\Sigma$. Consequently, it is tedious to propose a biophysic-matematical approach that involve time dependence of $\alpha$, $\beta$, $\phi_0$, $\phi_s$, $\eta_1$, $\eta_2$, $\varepsilon_1$, and $\varepsilon_2$ in time changes of $R_T$ and $\sigma_{12}$ at $\Sigma$. With this in mind, a longitudinal study is required to allow measuring each of these eight parameters in time. It is important to point out that values of $\phi_0$, $\phi_s$, $\eta_1$, $\eta_2$, $\varepsilon_1$, and $\varepsilon_2$ are reported in transversal studies [6,7,10-13,29,36,37,44]; therefore, these values cannot be extrapolated to other time instants.

Tumor necrosis is caused by nutrient and oxygen deprivation, and metabolic stress. The content of necrotic cells enhances angiogenesis and proliferation of endothelial cells, induces vasculature, as well as increases migration, invasion, and cell-cell interaction. Both necrosis and angiogenesis impact directly on cancer promotion and on the tumor microenvironment, as well as on cancer resistance and recurrence [56,57]. The influence of necrosis and angiogenesis on $\sigma_{12}$ at $\Sigma$ may be explained from equations (12) and (14). The tumor necrosis leads to an increase of $\alpha$ parameter, whereas tumor angiogenesis brings about an increase of the parameter $\alpha$ and a decrease of the parameter $\beta$ ($1/\beta$ dominates the term $(1 - e^{-\beta t})$ in the exponent of equation (14)). Consequently, $R_T$ increases and $\sigma_{12}$ at $\Sigma$ decreases in absolute value ($\sigma_{12}$ at $\Sigma$ makes more positive) in both cases. It should be noted that decrease of $\beta$ during tumor growth means that the balance between the productions of angiogenic and antiangiogenic molecules is dominated by angiogenic molecules.

The cell loss factors (CLFs: necrosis, apoptosis, exfoliation and metastasis) should be carefully analyzed in untreated tumors. These CLFs should be small so that the doubling time of the tumor (DT) be short, according to Steel equation ($DT = T_c \ln 2/[(1 - CLFs)(1 + GF)]$, where $T_c$ and GF are the cell cycle average time and tumor growth factor, respectively) [58]. For instance, our vast experience in preclinical studies indicate that the tumor necrosis percentage varies between 10-30 % of the entire tumor volume, depending on tumor histological variety, $V_{T0}$, host, and observation period of the study [1,2,4]. Short DT leads to an increase of $\alpha$ and decrease of $\sigma_{12}$ at $\Sigma$. This may be explained from the following expression obtained by substituting $V_T(t = DT) = 2V_{T0}$ in equation (13), given by $\alpha = \beta \ln 2/(1 - e^{-\beta t})$. An increase of number of cells that participate in cell cycle ($N_{cc}$) leads to an increase of GF ($GF = N_{cc}/(N_{cc} + N_{n-cc})$), where $N_{n-cc}$ is the number of cells that do not participate in cell cycle. As a result the increase of GF, DT is short, $\alpha$ increases and $\sigma_{12}$ at $\Sigma$ decreases.

When the tumor grows it becomes more heterogeneous, as it demonstrate simulations for with spherical and non-spherical geometries of it [43]. The tumor heterogeneity has one of main roles on cancer promotion and on the tumor microenvironment, as well as on cancer resistance and recurrence [8,14,39,43,56,57]. Therefore, it is considered a cancer hallmark. From the biological point of view, tumor more heterogeneous brings about an increase of $\alpha$ and therefore a decrease of $\sigma_{12}$ at $\Sigma$. This statement is corroborated from a bioelectric point of view with equation (12), as discussed above.

The simulations shown in [43] suggest that the spherical tumor has greatly defined its layers compared to ellipsoidal tumors, which may validate why the spherical tumor is a good model to study chemo-resistance and radio-resistance [41,42]. The tumor heterogeneity may be simulated approximately from a biophysical point of view following the same ideas of this study. Thus, we assume the spherical tumor formed by $M_T$ concentric layers, each one of them of radius $R_{Ti}$, average electrical conductivity $\eta_i$ and average electrical permittivity $\varepsilon_i$; -$\phi_0$ in the tumor center; -$\phi_{si}$ in the contour between two adjacent layers ($\Sigma_{i(i+1)}$) and it satisfies -$\phi_0 < -\phi_{s1} < \ldots < -\phi_{sN}$, keeping in mind [44]; and the existence of a surface charge density ($\sigma_{i(i+1)}$) at $\Sigma_{i(i+1)}$, such that: -$\sigma_{i(i+1)} < -\sigma_{(i+1)(i+2)}$ (i = 1,..., $M_T$). Furthermore, there is a surface charge density ($\sigma_{M(M+1)}$) at border ($\Sigma_{M(M+1)}$) between the outermost layer of the tumor and surrounding healthy tissue (average electrical conductivity $\eta_{M+1}$ and average electrical permittivity $\varepsilon_{M+1}$). For this case, results

$$\sigma_{i(i+1)} = -\varepsilon_{i+1}\left[\frac{\eta_i}{\eta_{i+1}} - \frac{\varepsilon_i}{\varepsilon_{i+1}}\right]\left[\frac{\phi_0 - \phi_{si}}{R_i}\right] , i = 1,..., M_T. \tag{15}$$

The condition -$\sigma_{i(i+1)} < -\sigma_{(i+1)(i+2)}$ (i = 1,..., $M_T$) supposes that each tumor layer behaves as an electrical shield, being the innermost layer the most negative, as discussed above. In contrast, the solid tumor is self-destructed, in contrast with the clinics [1-4]. Furthermore, the existence of -$\sigma_{i(i+1)}$ (i = 1,..., $M_T$) may explain that spherical tumor has well-defined multicentric layers from an electrical point of view, in agreement with well-defined multicentric layers from a biological point of view [43]. This confirms the close relationship between electrical and physiological parameters in biological tissues [10,11,14,15,36]. Nevertheless, the equation (15) has the inconvenient that $M_T$, $R_{Ti}$, $\eta_i$, $\varepsilon_i$, $\phi_{si}$, $\sigma_{i(i+1)}$ at $\Sigma_{i(i+1)}$, $\sigma_{M(M+1)}$ at $\Sigma_{M(M+1)}$ (i = 1,..., $M_T$) are not known neither experimentally nor theoretically. The measurement of these parameters in a multilayer tumor is more cumbersome than in a simple model, as the proposed in this study. That is why, we do not include a tumor with different concentric layer in the simulations proposed.

The electrical properties and active bioelectricity inherent in cancer and surrounding healthy tissue, as a whole, cannot be analyzed as the sum of all processes that occur at the molecular and cellular levels. This may be argued because biological systems are by nature multiscale and formed by closely interconnected and hierarchically organized multiple subsystems and supersystems, resulting in large networks of physical or functional proximities. Subsystems are referred to biological entities in the order of nanometers (i.e., amino acids residues), angstroms (i.e., single atoms), tents to hundreds of nanometers (i.e., proteins), several microns (i.e., organelles, cells). Supersystems are referred to tissues, organs and individuals measured in fractions of meters [59].

Large networks of systems in cancer patients allow to suggest that alterations in they are not only due to changes at tissue, cellular and molecular levels [8], but also to nanometric changes, as report in [60]. Furthermore, the integral characterization of cancer patients by means of an integrated analysis of clinical-biological(tumor and patient)-functional-bioelectrical parameters [61] is possible to these larger networks. The cancer fractality at submicron [60] and tissue [1,4,5] levels confirms the close relation of the multiscale hierarchies in malignant tumors.

## 4.1 Insights about cancer therapy

Many molecules used in chemotherapy and immunotherapy are positively/negatively charged and have not given a definitive solution to the cure of cancer. Our simulations indicate that anti-cancer therapies should take into account that bioelectricity cancer cells and $\sigma_{12}$ at $\Sigma$ are negative to reestablish the bioelectrical states and $V_{mem}$ of cancer cells within the physiological range, as report

Cervera et al. [9], who recommend that the use of non-physiological perturbations would not be necessary for cancer. Although the exact mechanism is poorly understood, different cancer types generate specific galvanotaxis responses to low direct current electric fields [8,62]. The results of this study confirm that anodes (positive electrodes) should be inserted in tumor regions near $\Sigma$ to avoid metastasis of cancer cells when electrochemical therapy is used. This may be explained because anodes generate positively charged carriers (e.g., hydrogen ions) that may intensify electrostatic interactions between negatively charged carriers (e.g., cancer cells) by means of the formation of ionic bridges. Consequently, a possible anti-cancer therapy that inhibits the release of positively charged carriers from the tumor may be suggested. Furthermore, knowledge of shape and orientation of $\sigma_{12}$ may be essential to elucidate if anodes should be inserted in regions near $\Sigma$ with higher or smaller $\sigma_{12}$ values to maximize tumor volume destruction with the minimum damage to the surrounding healthy tissue.

This study opens new questions that may be essential to understand TGK and how electrophysiological variables of the untreated tumor change during its growth that may be relevant for individualized anti-cancer therapies. Among the questions these arise. 1) Does the tumor growth bring about change from $\sigma_{12-0}$ to $\sigma_{12-f}$ at $\Sigma$ or does this change leads to the tumor growth? 2) Do biological changes (e.g., metabolism abnormalities) lead to physical changes (e.g., changes in $V_{mem}$ and electrical properties) [43] or vice versa [40]. We believe that dynamical bioelectrical changes are primary mechanisms involved in cancer that lead to chemical changes, to biological modifications, and to clinical alterations (secondary mechanisms). 3) Are the negative charged molecules across $\sigma_{12}$ at $\Sigma$ (from surrounding cancer tissue) more easily than the positive ones? 4) Do the negatively charged molecules that across $\sigma_{12}$ at $\Sigma$ induce the highest antitumor effectiveness than those positively charged? A meta-analysis may be carried out to give answer to these questions and others related to them. The cancer bioelectric handling has been suggested as a useful tool to understand bioelectric fields that change dynamically during cancer growth and possible anti-cancer therapeutic targets, aspects that remain unclear yet [8,43]. 5) What relationship exists between $\sigma_{12}$ and the tumor contour fractal dimension reported in [1,4,5]? 6) What implication non-homogeneous distribution of $\sigma_{12}$ at $\Sigma$ has during tumor growth? 7) What expression adopts $\sigma_{12}$ when a heterogeneous tumor and nonlinear $\phi_1$ are considered? 8) Can electrochemical therapy with low-level of direct current re-establish physiological bioelectrical disorders that happen in an untreated tumor? 9) How do the endogenous magnetic field and the ellipsoidal geometry influence the untreated tumor growth? 10) How does $\sigma_{12}$ relate to other biophysical-chemical processes that occur in the tumor? 11) How does $\sigma_{12}$ at $\Sigma$ change experimentally over time during the growth of untreated and treated solid tumors using any experimental techniques reported in [34,35] (e.g., electrostatic force microscopy)?

On the other hand, the results of this study may contribute to give answer to the fourth challenge reported in [59], related to the capture of dynamics in multiscale models because nanometric, atomic, molecular, cellular and tissue processes are highly dynamic [1,59].

## 5 Conclusion

In conclusion, graphic strategies corroborate the correspondence between the electrical and physiological parameters in the untreated cancer, which may have an essential role in its growth, progression, metastasis and protection against immune system attack and anti-cancer therapies. In addition, knowledge of $\sigma_{12}$ at $\Sigma$ may be relevant in the redesign of chemotherapy and immunotherapy that have into account the polarity of the substances or in the design of completely novel therapies.

## Acknowledgements

The authors would like to thank two unknown reviewers, Dr. Drasdo (Associate Editor), Dr. Cicuta (Subject Editor) and PhD. Hall (Editor-in-Chief) for their valuable comments and suggestions that greatly improved this manuscript. All these professionals belong to the editorial board of the Royal Society Open Science.

## Author Contributions

**H.B.P.:** Organization and Writing-original draft, Contributed to physics-mathematical models, Performed the simulations, Revised the manuscript and Supervised the entire work; **A.A.S.J.:** Organization and Writing-original draft, Revised the manuscript and Supervised the entire work; **E.J.R.O.:** Conceptualization, Organization and Writing-original draft, Contributed to physics-mathematical models, Revised the manuscript and Supervised the entire work; **J.A.H.K.:** Contributed to physics-mathematical models; **M.M.G.:** Conceptualization, Organization and Writing-original draft; **N.A.V.G.:** Conceptualization and Writing-original partial draft. **N.H.M.:** Organization and Writing-original draft; **V.G.S.G.:** Organization and Writing-original draft; **J.I.M.T.:** Conceptualization, Organization and Writing-original draft, Contributed to physics-mathematical models, Performed the simulations, Revised the manuscript and Supervised the entire work; **L.E.B.C.:** Conceptualization, Organization and Writing-original draft, Contributed to physics-mathematical models, Performed the simulations, Revised the manuscript and Supervised the entire work.

## Declaration of Competing Interests

We have no competing interests.

## Funding

This research is partially supported by CITMA from Santiago de Cuba, Cuba [Project PT241SC003-002]; and Ministerio de Ciencia e Innovación, Spain [Project PID2019-109045GB-C31].